\documentclass[12pt]{article}
\usepackage{pic03}
\usepackage{hyperref}
\usepackage{url}
\usepackage{graphicx}

\begin{document}

\title{\bf Muon $g-2$}
\author{
Ernst Sichtermann, for the Muon $g-2$ collaboration\,\cite{collaboration}  \\
{\em Yale University -- Department of Physics, New Haven, Connecticut 06520}}
\maketitle

%
%
%
%
%
%
\vspace{4.5cm}
%

\baselineskip=14.5pt
\begin{abstract}
The Muon $g-2$ collaboration has measured the anomalous magnetic $g$ value
of the positive muon to within a relative uncertainty of 0.7\,parts per million.
The result, \mbox{$a_{\mu^+}(\mathrm{expt}) = 11\,659\,204(7)(5)\ \times 10^{-10}$},
is in good agreement with the preceding data on $a_{\mu^+}$ and $a_{\mu^-}$ and
has about twice smaller uncertainty.
The measurement tests standard model theory, which at the level of the present
experimental uncertainty involves quantum electrodynamics, quantum
chromodynamics, and electroweak interaction in significant ways.
The analysis of the data on the anomalous magnetic $g$ value of the
negative muon is well underway.
\end{abstract}
\newpage

\baselineskip=17pt

\section{Introduction}
The anomalous $g$ values, $a = (g-2)/2$, of leptons arise from quantum
mechanical effects.
Their precise measurement has historically played an important role in the
development of particle theory. 
The anomalous magnetic $g$ value of the electron, $a_e$,
has been measured to within about four parts per billion (ppb)~\cite{VanDyck:1987ay},
and is thus among the most accurately known quantities in physics.
Its value is described in terms of standard model (SM) field
interactions, with nearly all of the measured value contributed by QED
processes involving virtual photons, electrons, and positrons.
Heavier particles contribute to $a_e$ only at the level of
the present experimental uncertainty.

The anomalous magnetic $g$ value of the muon, $a_\mu$, is more sensitive than
$a_e$ to processes involving particles more massive than the electron,
typically by a factor $(m_\mu/m_e)^2 \sim 4 \cdot 10^4.$
A series of three experiments~\cite{cern-123} at CERN measured
$a_\mu$ to within 7~parts per million (ppm), which is predominantly statistical.
The CERN generation of experiments thus tested electron-muon universality
and established the existence of a hadronic contribution to $a_\mu$ with a relative
size of $\sim 59$\,ppm.
Electroweak processes are expected to contribute at the level of 1.3\,ppm, as are
many speculative extensions of the SM.

The muon $g-2$ experiment at Brookhaven National Laboratory (BNL)
is conceptually similar to the last CERN experiment, and has determined
$a_{\mu^+}$ of the positive muon with an uncertainty of 0.7\,ppm from a sample
of about $4 \cdot 10^9$ decay positrons collected in the year 2000.
The analysis of  $a_{\mu^-}$ of the negative muon from a similarly sized sample
of decay electrons collected in 2001 is well underway.

\section{Experiment}
\begin{figure}
  \begin{center}
  \includegraphics[width=0.6\textwidth]{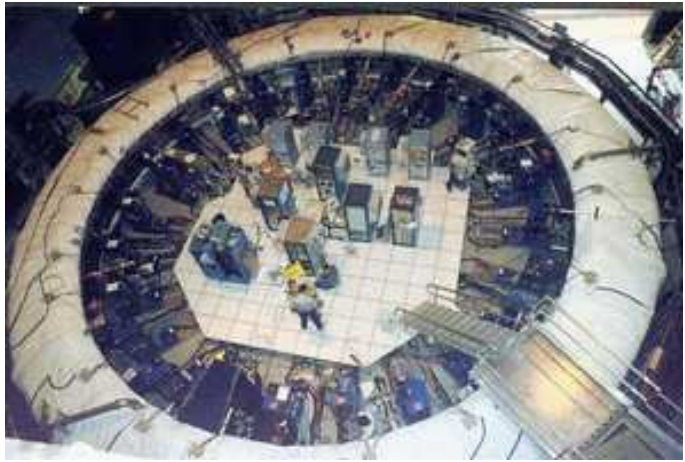}
  \end{center}
  \caption{Top view of the $g-2$ apparatus.  The beam of longitudinally polarized muons enters the superferric storage ring magnet through a superconducting inflector magnet located at 9 o'clock and circulates clockwise after being placed onto stored orbit with three pulsed kickers modules in the 12 o'clock region.  Twenty-four lead scintillating-fiber calorimeters on the inner, open side of the C-shaped ring magnet are used to measure muon decay positrons (electrons).
The central platform supports the power supplies for the four electrostatic quadrupoles and the kicker modules.}
  \label{fig:apparatus}
\end{figure}

The measurements at the Brookhaven Alternating Gradient Synchrotron used a secondary
pion beamline.
For most of the data taking periods, longitudinally polarized muons of about 3.1\,GeV from
forward decays were momentum-selected and injected into a 14.2\,m diameter storage ring
magnet~\cite{magnet} through a field-free inflector~\cite{inflector} region in the magnet yoke.
A pulsed magnetic kicker~\cite{kicker} located at approximately one quarter
turn from the inflector region produced a 10\,mrad deflection which placed
the muons onto stored orbits.
Pulsed electrostatic quadrupoles~\cite{quads} provided vertical focusing.
The magnetic dipole field of about 1.45\,T was measured with an NMR
system~\cite{nmr} relative to the free proton NMR frequency $\omega_p$
over most of the 9\,cm diameter circular storage aperture.
Twenty-four electromagnetic calorimeters~\cite{calo} read out by
400\,MHz custom waveform digitizers (WFD) were used on the open, inner
side of the C-shaped ring magnet to measure the decay positrons and electrons.
Muon decay violates parity, which in the laboratory frame results in a
modulation of the number of positrons (electrons),
\begin{equation}
  N(t) = N_0(E) \exp\left(\frac{-t}{\gamma\tau}\right)
       \left[ 1 + A(E) \sin \left( \omega_a t + \phi(E) \right) \right],
  \label{eq:precession}
\end{equation}
above an energy threshold $E$.  Here, $N_0$ is a normalization,
$\gamma\tau \sim 64\,\mu\mathrm{s}$ is the dilated muon lifetime,
$A \sim 0.4$ is an asymmetry factor, $\phi$ is a phase, and $\omega_a$ is
the angular difference frequency of muon spin precession and
momentum rotation.

The muon anomalous magnetic $g$ value is evaluated from the ratio of the
measured frequencies, $R = \omega_a/\omega_p$, according to:
\begin{equation}
  a_\mu = \frac{R}{\lambda - R},
\end{equation}
in which $\lambda = \mu_\mu / \mu_p$ is the ratio the muon and proton
magnetic moments.
The value with smallest stated uncertainty,
$\lambda = \mu_\mu / \mu_p =  3.183\,345\,39(10)$~\cite{pdg},
results from measurements of the microwave spectrum of ground state
muonium~\cite{liu} and theory~\cite{kino}.

\section{Data Analysis}
The proton NMR frequency $\omega_p$ and the muon spin precession
frequency $\omega_a$ were analyzed independently by several groups
within the collaboration.
The values of $R = \omega_a/\omega_p$ and $a_\mu$ were evaluated only
after each of the frequency analyses had been finalized; at no earlier
stage were the absolute values of both frequencies known to any of the collaborators.

\subsection{The magnetic field}
\begin{figure}
  \includegraphics[width=0.95\textwidth]{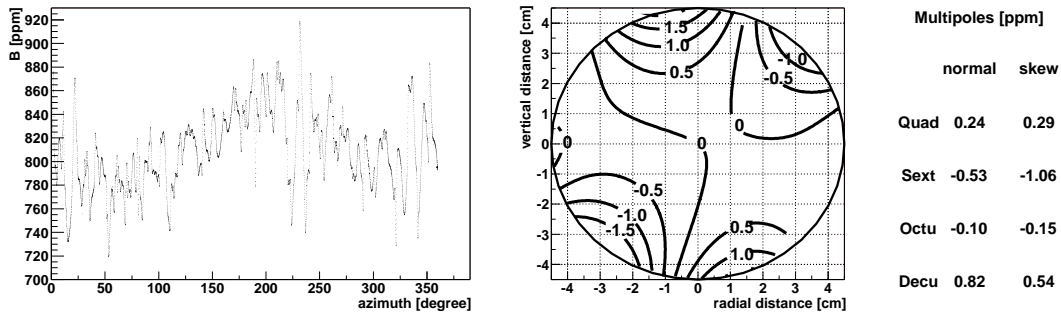}
  \caption{The NMR frequency measured with the center trolley probe relative
  to a 61.74\,MHz reference versus the azimuthal position in the storage ring
  (left), and (right) a 2-dimensional multipole expansion of the azimuthal
  average of the field measured with 15 trolley probes with respect to the
  central field value of 1.451\,275\,T.
  The multipole amplitudes are given at the storage ring aperture, which has
  a 4.5\,cm radius as indicated by the circle.}
  \label{fig:field}
\end{figure}

During the data collection period from January to March 2000, a field trolley with
17 NMR probes was used 2-3 times per week, 22 times in total, to measure the
field throughout the muon storage region.
Figure~\ref{fig:field}a shows the field value measured in the storage ring with
the center trolley probe versus the azimuthal angle.
The field is seen to be uniform to within about $\pm 50\,$ppm of its average
value over the full azimuthal range, in particular also in the  region near
$350^\circ$ where the inflector magnet is located.
Since the field averaged over azimuth is uniform to within 1.5\,ppm over the storage
aperture (Fig.~\ref{fig:field}b), the field integral encountered by the (analyzed)
muons is rather insensitive to the precise location and profile of the beam.

The  probes inside the field trolley were calibrated with respect to each other
 during the data collection period using dedicated measurements in
which a single NMR probe was plunged into the storage vacuum.
This so-called plunging probe, as well as a subset of the trolley probes, were
calibrated {\em in situ} with respect to a standard probe~\cite{fei}.

The 22 measurements with the field trolley were used to relate the readings
of 370 NMR fixed probes in the outer top and bottom walls of the storage
vacuum chamber to the field values in the beam region, so that the fixed
NMR probe readings could be used to interpolate the field when the field
trolley was 'parked' in the storage vacuum just outside the beam region
and muons circulated in the storage ring.

For the data collection between January and March 2000,
the field frequency $\omega_p$ weighted by the muon distribution was
found to be,
\begin{equation}
  \omega_p/(2\pi) = 61\,791\,595(15)\,\mathrm{Hz}~~\mbox{(0.2\,ppm)},
  \label{eq:omegap_result}
\end{equation}
where the uncertainty has a leading contribution from the calibration of
the trolley probes and is thus predominantly systematic.
The result was confirmed by a second, largely independent analysis, which
made use of additional calibration data, a different selection of fixed NMR
probes, and a different method to relate the trolley and fixed probe readings.
The result from the data collection on the negative muon in the year 2001 is
expected to have further improved uncertainty.

\subsection{The muon spin precession frequency}
\begin{figure}
  \includegraphics[width=0.95\textwidth]{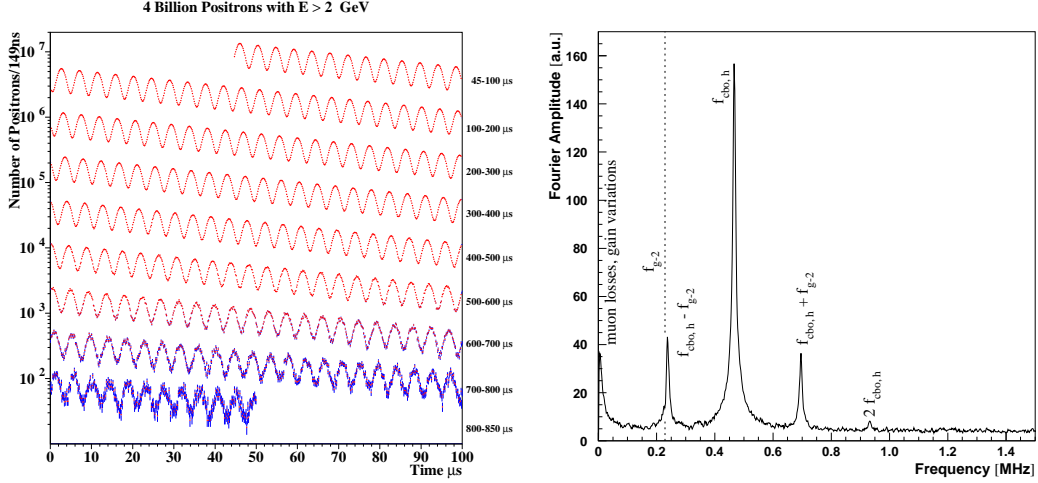}
  \caption{The time spectrum for $4 \cdot 10^9$ positrons with energies greater than 2\,GeV collected from January to March 2000, after corrections for pile-up and for the bunched time structure of the injected beam (left) were made, and (right) the Fourier transform of the time spectrum, in which muon decay and spin precession (cf. Eq.~2) has been suppressed to emphasize other effects.
  }
  \label{fig:precession}
\end{figure}

About $4\cdot10^9$ reconstructed positrons  with energies greater than 2\,GeV and times  between  \mbox{50\,$\mu$s} and \mbox{600\,$\mu$s} following the beam injection were available
for analysis from the data collection between January and March 2000.
Figure~\ref{fig:precession}a shows their time spectrum after corrections for
the bunched time structure of the beam and for overlapping calorimeter
pulses, so called pile-up~\cite{Brown:2001mg}, had been applied.

The main characteristics of the spectrum are muon decay and spin
precession (Eq.~\ref{eq:precession}), however, additional effects need
to be considered as illustrated by the Fourier spectrum in Fig.~\ref{fig:precession}b.
These effects include detector gain and time instability, muon losses,
and oscillations of the beam as a whole, so-called coherent betatron
oscillations (CBO).
The latter are caused by the injection of the beam through the relatively narrow
$18(\rm{w}) \times 57(\rm{h})\,\rm{mm}^2$ aperture of the 1.7\,m long
inflector channel into the 90\,mm diameter aperture of the storage region.
Their frequencies are determined by the focusing index of the storage ring, and
have been observed directly with fiber harp monitors that were plunged into the
beam region for this purpose.
Since the calorimeter acceptances vary with the muon decay position in
the storage ring and with the momentum of the decay positron, CBO cause
modulation of the time and energy spectra of the observed positrons.

Numerically most important to the determination of $\omega_a$ from the
data collected between January and March 2000 were the CBO in the
horizontal plane, whose frequency was numerically close to twice the
frequency $\omega_a$.
When the modulations of the asymmetry and phase with frequency
$\omega_\mathrm{cbo,h} \simeq 2 \times \omega_a$ were not accounted
for in the function fitted to the data, artificial shifts of up to 4\,ppm in the
frequency values $\omega_a$ determined from individual calorimeter
spectra were observed.
In the joined calorimeter spectrum, such shifts are largely canceled
because of the circular symmetry of the experiment design.

Several approaches were pursued in the analysis of $\omega_a$.
In one approach, the time spectra from individual positron calorimeters
was fitted in narrow energy intervals using a fit function as in
Eq.~\ref{eq:precession} extended by the aforementioned
number, asymmetry, and phase modulations.
Other approaches made use of the cancellation in the joined calorimeter
spectra and either fitted for the residual of the leading effects, or
accounted for their neglect in a contribution to the systematic uncertainty.
The results were found to agree, on $\omega_a$ to within the expected 0.5\,ppm
statistical variation resulting from the slightly different
selection and treatment of the data in the respective analyses.
The combined result was found to be,
\begin{equation}
  \omega_a/(2\pi) = 229\,074\,11(14)(7)\,\mathrm{Hz}~~\mbox{(0.7\,ppm)},
\end{equation}
in which the first uncertainty is statistical and the second is systematic.
The above frequency includes a correction of +0.76(3)\,ppm for the net contribution to
the muon spin precession and momentum rotation caused by vertical beam
oscillations and, for muons with $\gamma \neq 29.3$, by horizontal electric fields~\cite{farley}.
The systematic uncertainty has a leading contribution of 0.2\,ppm caused by CBO.
In the year 2001, an event sample of comparable size on the negative muon was collected. 
The storage ring was operated with two different values of the focusing index, which
is expected to reduce the leading systematic uncertainty.

\section{Results and Discussion}
\begin{figure}
  \begin{center}
  \includegraphics[width=0.60\textwidth]{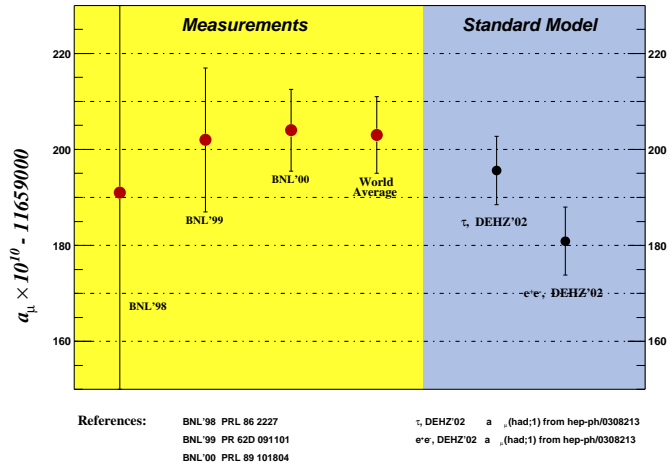}
  \end{center}
  \caption{Recent measuremens of $a_\mu$ and standard model evaluations
  using the estimates in Ref.~\cite{Davier:2003} of the lowest order
  contribution from hadronic vacuum polarization.}
  \label{fig:result}
\end{figure}

The value of $a_\mu$ was evaluated after the analyses of
$\omega_p$ and $\omega_a$ had been finalized,
\begin{equation}
  a_{\mu^+} = 11\,659\,204(7)(5)\ \times\ 10^{-10}~~\mbox{(0.7\,ppm)},
\end{equation}
where the first uncertainty is statistical and the second systematic.
The result agrees well with the preceding
measurements~\cite{cern-123,Brown:2001mg,Brown:2000sj} and
drives the present world average,
\begin{equation}
  a_\mu(\mathrm{exp}) = 11\,659\,203(8)\ \times\ 10^{-10}~~\mbox{(0.7\,ppm)},
\end{equation}
in which the uncertainty accounts for known correlations between the
systematic uncertainties in the measurements.
Figure~\ref{fig:result} shows our recent measurements of $a_{\mu^+}$,
together with two SM evaluations discussed below.

In the SM, the value of $a_\mu$ receives contributions
from QED, hadronic, and electroweak processes,
$a_\mu(\mathrm{SM}) =
  a_\mu(\mathrm{QED})  + a_\mu(\mathrm{had}) + a_\mu(\mathrm{weak})$.
The QED and weak contributions can, unlike the hadronic contribution, be
evaluated perturbatively,
$a_\mu(\mathrm{QED}) = 11\,658\,470.57(29)\,\times\,10^{-10}$~\cite{Mohr:2000ie}
and $a_\mu(\mathrm{weak}) = 15.4(2)\,\times\,10^{-10}$~\cite{Czarnecki:2002nt}.
The hadronic contribution is, in lowest order, related by dispersion theory
to the hadron production cross sections measured in $e^+e^-$ collisions
and, under additional assumptions, to hadronic $\tau$-decay.
Clearly, the hadronic contribution has a long history of values
as new data appeared and analyses were refined.

At the time of the PIC-2003 conference, a recent and detailed evaluation was the one
by Davier and co-workers, which -- unlike preceding analyses -- incorporated the low-energy
$e^+e^-$ annihilation cross section into hadrons by the CMD-2 collaboration~\cite{Akhmetshin:2001ig} , $e^+e^-$ measurements~\cite{Bai:1999pk} with improved accuracy in the
2--5\,GeV energy region from BES, preliminary results from the final ALEPH analysis~\cite{aleph} of hadronic $\tau$-decay at LEP1, and additional data~\cite{Anderson:1999ui} from CLEO.
Significant discrepancies between the $e^+e^-$ and $\tau$ data were found.

The CMD-2 collaboration has since released a reanalysis of their cross section measurements~\cite{Akhmetshin:2003} and Davier and co-workers have provided updated
estimates for the contribution to $a_\mu(\mathrm{SM})$ from lowest order hadronic
vacuum polarization,
$a_\mu(\mathrm{had,1}) = 696(7)\,\times\,10^{-10}$ from $e^+e^-$ data and
$a_\mu(\mathrm{had,1}) = 711(6)\,\times\,10^{-10}$ from $\tau$ data~\cite{Davier:2003}.
The authors refrain from averaging the values, noting that significant
discrepancies remain in the underlying data in the center-of-mass region between
0.85 and 1.0\,GeV.
Radiative return measurements at the $e^+e^-$ factories may reach the required precision to shed light on the situation, as might lattice calculations.
Higher order contributions to $a_\mu(\mathrm{had})$ include higher order hadronic vacuum
polarization~\cite{Krause:1997rf} and hadronic light-by-light scattering~\cite{Hayakawa:2001bb}.

Open questions thus concern the SM value of $a_\mu$, in particular its hadronic contribution,
and the experimental value of $a_{\mu^-}$ at sub-ppm precision.  Stay tuned!

\section{Acknowledgements}
This work was supported in part by the U.S. Department of Energy, the U.S. National Science Foundation, the U.S. National Computational Science Alliance, the German Bundesminister f\"{u}r Bildung und Forschung, the Russian Ministry of Science, and the U.S.-Japan Agreement in High Energy Physics.


\end{document}